\documentclass[review,2p]{elsarticle}

\journal{Computer Physics Communications}

\usepackage[a4paper]{geometry}
\usepackage[utf8]{inputenc}
\usepackage{amsmath}
\usepackage{bm}
\usepackage{MnSymbol}
\usepackage{graphicx}

\usepackage{tikz}
\usetikzlibrary{shapes,arrows,positioning,calc,backgrounds}

\def\qprop{\textsc{Qprop}}
\def\eulere{\mathrm{e}}
\def\imagi{\mathrm{i}}
\def\grad{\bm{\nabla}}

\def\abar{{\bar a}}
\def\Ibar{{\bar I}}

\def\halb{\frac{1}{2}}

\def\diff{\mathrm{d}}
\def\pabl#1#2{\frac{\partial #1}{\partial #2}}
\def\abl#1#2{\frac{\diff #1}{\diff #2}}

\def\bra#1{\langle #1 \vert}
\def\braket#1#2{\langle #1 \vert #2 \rangle}
\def\expect#1#2{\langle #1 \vert #2 \vert #1 \rangle}
\def\ket#1{\vert #1 \rangle}

\def\vece{\mathbf{e}}
\def\vecr{\mathbf{r}}
\def\veck{\mathbf{k}}
\def\vecp{\mathbf{p}}
\def\vecA{\mathbf{A}}
\def\vecE{\mathbf{E}}
\def\vecalpha{\bm{\alpha}}

\def\unitvec#1{\mathbf{e}_{#1}}

\def\RI{{R_{\text{I}}}}
\def\TP{T_{\text{p}}}
\def\XI{X_{\text{I}}}
\def\UP{U_{\text{p}}}
\def\varphicep{\varphi_{\mathrm{CEP}}}

\begin{document}
\begin{frontmatter}

\title{Photoelectron spectra with \qprop{} and t-SURFF}

\author{Volker Mosert}
\author{Dieter Bauer\corref{mycorrespondingauthor}}
\address{Institut für Physik, Universität Rostock, 18051 Rostock, Germany}
\cortext[mycorrespondingauthor]{Corresponding author}
\ead{dieter.bauer@uni-rostock.de}



\begin{abstract}
Calculating strong-field, momentum-resolved photoelectron spectra (PES) from numerical solutions of the time-dependent Schr\"odinger equation  (TDSE) is a very demanding task due to the large spatial excursions and drifts of electrons in intense laser fields.
The time-dependent surface flux (t-SURFF) method for the calculation of PES [L.\ Tao, A.\ Scrinzi, New Journal of Physics 14, 013021 (2012)] allows to keep the numerical grid much smaller than the space over which the wavefunction would be spread at the end of the laser pulse. 
We present an implementation of the t-SURFF method in the well established TDSE-solver \qprop{}
[D.\ Bauer, P.\ Koval, Comput.\ Phys.\ Commun.\ 174, 396 (2006)].
\qprop{} efficiently propagates wavefunctions for single-active electron systems with spherically symmetric binding potentials in classical, linearly (along $z$) or elliptically (in the $xy$-plane) polarized laser fields in dipole approximation. Its combination with t-SURFF makes the simulation of PES feasible in cases where it is just too expensive to keep the entire wavefunction on the numerical grid, e.g., in the long-wavelength or long-pulse regime.
\end{abstract}


\end{frontmatter}

{\bf NEW VERSION PROGRAM SUMMARY}

\begin{small}
\noindent
{\em Manuscript Title:} Photoelectron spectra with \qprop{} and t-SURFF \\
{\em Authors:} Volker Mosert, Dieter Bauer \\
{\em Program Title:} \qprop{} \\
{\em Journal Reference:} \\
{\em Catalogue identifier:}                                   \\
{\em Licensing provisions:} GNU General Public License 3 (GPL) \\
{\em Programming language:}
C++                                   \\
{\em Computer:} x86\_64                                              \\
{\em Operating system:}
Linux                                       \\
{\em RAM:} The memory requirements for calculating PES are determined by the maximum $\ell$ in the spherical harmonics
expansion of the wave function and the number of momentum (or energy) values for which the PES is to be calculated. The
example with the largest memory demand ({\tt large-clubs}) uses
approximately 6GB of RAM.
The size of the numerical representation of a wavefunction during
propagation is modest for the examples included (53 MB for the
{\tt large-club} example).
{\em Number of processors used:} The evaluation of the PES can be distributed over up to $N_k$ MPI processes ($N_k$ is the number of momentum values). \\
{\em Supplementary material:}  See {\tt www.qprop.de}.                               
{\em Classification:} 2.5 photon interactions                                         \\
{\em External routines/libraries:} GNU Scientific Library, Open MPI (optional), BOOST (optional) \\
{\em Subprograms used:}                                       \\
{\em Catalogue identifier of previous version:} ADXB\_v1\_0 \\
{\em Journal reference of previous version:} Comput. Phys. Commun. 174 (2006) 396. \\
{\em Does the new version supersede the previous version?:} For TDDFT calculations the previous version should be used. \\
{\em Nature of problem:} When atoms are ionized by intense laser fields electrons may escape with large momenta (especially when rescattering is involved).
This translates to a rapidly spreading wavefunction in numerical simulations of these systems 
thus rendering the calculation of PES  very costly for increasing wave lengths and peak intensities.  \\
{\em Solution method:} The TDSE is solved by propagating the wavefunction using a Crank-Nicolson propagator.
The wavefunction is represented by an expansion in spherical harmonics.
In order to reduce the requirements with respect to the grid size the t-SURFF method is used to calculate PES. \\
{\em Reasons for the new version:} Using the window operator method to calculate PES is increasingly costly with increasing  ponderomotive energies and pulse durations.
The new version of \qprop{} provides an implementation of the t-SURFF method which allows the use of much smaller numerical grids.\\
{\em Summary of revisions:} An implementation of the t-SURFF method and examples for calculating PES  are provided in the new release.\\
{\em Restrictions:} The dipole approximation for the laser interaction has to be applicable.
 t-SURFF is only implemented for velocity gauge.
Furthermore a finite cutoff for long range binding potentials has to be used in the implemented t-SURFF method.\\
{\em Running time:} Depends strongly on the laser interaction studied. The examples given in this paper have run times from a few minutes to 12.5 hours. \\
   \\

\end{small}

\section{Introduction}\label{sec:intro}
Ten years after the initial release of \qprop{}~\cite{bauer2006qprop}, solving the TDSE 
for more than one ``active'' electron in strong laser fields remains a Herculean effort.
Consequently, single-active-electron TDSE simulations remain one of the most important tools in strong field physics. Yet, what seems just an innocent extension of the 111-year-old photoeffect, namely a single, initially bound electron in an intense laser field,  shows a plethora of unexpected features \cite{huismans2011,PhysRevX.5.021034} that still challenge theory to date.

One of the most demanding tasks in strong-field TDSE simulations is the calculation of momentum-resolved PES.
Especially for long-wavelength lasers the large final momenta and long pulse duration make large computational grids necessary, at least if the calculation of PES relies on the wavefunction after the laser pulse.
Employing wavelengths larger than the standard 800~nm might be beneficial for ``self-imaging'' using the target's own electrons \cite{cdlin2010} since the higher ponderomotive energies and return momenta will better probe the target structure without the need to increase the laser intensity towards the destructible over-barrier regime.

According to textbook quantum mechanics, momentum-resolved PES should be calculated as $|\braket{\phi_\vecp }{ \Psi(t\to\infty)}|^2$ where $\ket{\Psi(t\to\infty)}$ is the electronic state as $t\to\infty$ when the laser is off, and $\ket{\phi_\vecp }$ is a continuum eigenstate of asymptotic momentum $\vecp$. Not only is none of the assumptions true on a numerical grid, but also the eigenstates $\ket{\phi_\vecp }$ on the grid are unknown and expensive to compute for all $\vecp$ of interest (note that even if ${\phi_\vecp }(\vecr)$ is known analytically its self-consistent, discretized representation on the numerical grid is not). In efficient numerical calculations of approximate PES one typically works around projections on unperturbed eigenstates. 
Commonly, such methods are based on an approximate projection operator applied to the (in some way or another discretized) wavefunction immediately after the pulse (e.g., the ``window operator"~\cite{schafkul1990,schafer1991energy}), Fourier-transforms, spectral analysis in time (i.e., further unperturbed time-propagation and analysis of the autocorrelation function \cite{feitflecksteiger1982}), or so-called ``virtual detectors'' \cite{feuerthumm2003}. The t-SURFF method was proposed in Ref.~\cite{tsurff}, noticing that the spatially very extended wavefunction after the pulse may be traded for temporal information about the wavefunction on a surface enclosing a much smaller, central part of the grid. Surface-flux methods have been long well-known for time-independent Hamiltonians. The related problem of perfectly transparent boundary conditions in time-dependent calculations has been addressed as well \cite{bouckeschmitzkull1997,PhysRevE.88.053308}. The very significant achievement in Ref.~\cite{tsurff} is to employ the surface flux through a boundary while the laser is on for the calculation of PES.

TDSE solutions serve as  an important  benchmark for simpler, almost-analytical approaches such as the strong-field approximation and its quantum orbit flavors (see \cite{sergeyreview2014} for a recent review). 
The TDSE-solver \qprop{} enables the efficient simulation  of a single active electron, initially bound in a spherically symmetric potential and interacting with an intense laser field. 
In  \qprop{}, the  wavefunction is expanded in spherical harmonics $Y_{\ell m}(\Omega)$, and the radial wavefunctions $\phi_{\ell m}(r)$ are propagated in time. The laser field is treated in dipole approximation. For laser fields linearly polarized in $z$-direction, the orbital angular momentum component $\hat L_z$ is a constant of the motion, and the related magnetic quantum number $m$ remains ``good.'' As a consequence, the problem is effectively two-dimensional, and the partial waves are propagated on an $r\ell$-grid, where $r$ is the radial coordinate, and $\ell$ the orbital angular momentum quantum number. For linear polarization, the Muller algorithm introduced in~\cite{muller1999efficient} is used in \qprop. The Muller algorithm is based on the unconditionally stable and unitary Crank-Nicolson propagation with improved spatial discretizations in $r$ in combination with a decomposition in $2\times 2$ matrices acting in angular-momentum space.

\qprop{} is also able to propagate wavefunctions for arbitrary elliptical polarization in the $xy$-plane. However, owing to the broken azimuthal symmetry $m$ is not a good quantum number in this case, the problem is really three-dimensional, and the computational cost higher. The details of the propagation routines for linearly and elliptically polarized laser fields (beyond what is explained in \cite{bauer2006qprop}) can be found on the \qprop\ website \cite{qpropweb}. There is also a list of papers in which \qprop\ was used. The current paper only concerns the implementation of t-SURFF in  \qprop{}.

The original  \qprop{} package incorporates the possibility to perform time-dependent density functional (TDDFT) calculations, i.e., the solution of the time-dependent Kohn-Sham (KS) equations~\cite{Ullrich_11}. 
How to calculate {\em rigorously} many-electron PES from KS orbitals is an open question in TDDFT, interesting in itself~\cite{tddftpes2012} but not the topic of this work. 

Another restriction concerns the gauge.  \qprop{} allows for velocity and length gauge, and it has been thoroughly tested that observables converge to the same result. However, the computational cost in velocity gauge is much smaller for strong laser fields. This is  because---given a vector potential $\vecA(t)$---the large, purely oscillatory component $\sim \vecA(t)$ in the kinetic momentum $\vecp_{\mathrm{kin}}$ is absent in the canonical momentum $\vecp=\vecp_{\mathrm{kin}}-\vecA(t)$, allowing for smaller $\ell$-grids, larger time steps, and grid spacings \cite{optimal-gauge}.  On the other hand, problems where the binding potential around the origin dominates the energy scale are better treated in length gauge. Since t-SURFF is most beneficial to intense-laser problems we implemented it for the velocity gauge.

The paper is organized as follows: In  Section~\ref{sec:theory}, the elementary t-SURFF idea is reviewed, and the particular \qprop{} aspects are discussed. We occasionally refer to the code version discussed in this paper as \qprop~2.0. The main, important changes compared to the original Qprop version \cite{bauer2006qprop} are described  in Section~\ref{sec:lib} and summarized in Table~\ref{tab:changes}.

Examples for the calculation of momentum-resolved spectra for above-threshold ionization (ATI) in the multiphoton regime, a linearly polarized mid infrared laser pulse, and a circularly polarized few cycle laser pulse are provided in Section~\ref{sec:examples}. Atomic units $\hbar=m_{{e}}=\vert e \vert = 4\pi\epsilon_0 = 1$ are used throughout the paper except where indicated otherwise.

\section{Theory}\label{sec:theory}
The method for solving the TDSE used in \qprop{} is covered extensively in the original article~\cite{bauer2006qprop} and a technical manuscript available for download on the \qprop{} website~\cite{qpropweb}.
Hence, only the basic ideas will be summarized, and the focus clearly lies on the incorporation of t-SURFF in \qprop.
\subsection{Propagation of the wavefunction}\label{sec:propagation}
\qprop{} is applicable to systems with a single (active) electron, spherically symmetric binding potentials, and laser fields that can be described classically and in dipole approximation. The Hamiltonian may then be chosen as 
\begin{equation}
  \label{eq:hamiltonian-3D}
  \hat H=-\frac12\nabla^2 -\imagi\vecA(t)\cdot \grad + V(r) .
\end{equation}
Here, the velocity gauge is used, with the purely time-dependent $A^2(t)$ term already transformed away (see Appendix A).
In the absence of an external field $\vecA(t)$, spherical symmetry allows to separate the problem into uncoupled, one-dimensional Schr\"odinger equations for the radial wavefunctions $ \phi_{\ell m}(r,t)$ if the total wavefunction is expanded in spherical harmonics,
\begin{equation}
  \label{eq:wave-fun-expansion}
  \Psi(\vecr, t)=\frac{1}{r}\sum_{\ell = 0}^\infty \sum_{m = -\ell}^\ell \phi_{\ell m}(r,t) Y_{\ell m}(\Omega).
\end{equation}
Computationally, the upper limit for $\ell$ is finite, say $ L_{\max}-1$. We store the radial wavefunctions $\phi_{\ell m}(r,t)$ for $\ell=0,1,2, \ldots L_{\max}-1$, $m=-\ell, -\ell+1, \ldots, \ell$ on a uniformly discretized radial grid $r=i\Delta r$, $i=1,2, \ldots N_r$.
The propagation routine in \qprop{} supports two basic modes of operation. 
The first of these modes is designed for simulating the interaction with a  linearly (in $z$-direction) polarized laser field, $\vecA(t)=A_z(t)\vece_z$. In this case the magnetic quantum number $m$ of the initial state is conserved, $m=m_0$, 
\begin{equation}
  \label{eq:wave-fun-expansion_linpol}
  \Psi_{m_0}(\vecr, t)=\frac{1}{r}\sum_{\ell = 0}^{L_{\max}-1} \phi_{\ell}(r,t) Y_{\ell m_0}(\Omega) \qquad \mathrm{(lin.\ pol.)},
\end{equation}
and the set of radial wavefunctions $\{ \phi_{\ell}(r,t) \}$ are effectively propagated on a two-dimensional $r\ell$-grid. Here, we assume that, for simplicity, there is only one $m=m_0$ and not a superposition of various partial waves of different $m$. In the latter case, one simply may propagate the different $m$-components independently from one another.

The second  mode supports a laser field of arbitrary, i.e., elliptical, polarization in the $xy$-plane, $\vecA(t)=A_x(t)\vece_x + A_y(t)\vece_y$. In that case a full expansion, including all $m$s, is required,
\begin{equation}
  \label{eq:wave-fun-expansion_elliptpol}
  \Psi(\vecr, t)=\frac{1}{r}\sum_{\ell = 0}^{L_{\max}-1}  \sum_{m = -\ell}^\ell\phi_{\ell m}(r,t) Y_{\ell m}(\Omega) \qquad \mathrm{(ell.\  pol.)}.
\end{equation}
Whereas for linear polarization the run time scales $\sim L_{\max}$ it grows $\sim L_{\max}^2$ for elliptical polarization.

\subsection{Window operator method for photoelectron spectra}\label{sec:winop}
The window operator method (WOM) \cite{schafkul1990,schafer1991energy} is an efficient way to calculate PES if the complete wavefunction  after the interaction with the external field is available. Strictly speaking, already the initial eigenstate wavefunction in realistic binding potentials is nonzero everywhere (apart from  nodal planes, lines, or points). Computationally, it should be negligibly small on the boundary of the numerical grid. During the interaction with the external field, outgoing flux is typically removed by mask functions, absorbing  potentials \cite{PhysRevA.74.034701}, or complex scaling \cite{PhysRevA.81.053845}. The parts of the wavefunction that have been removed in such a way are lost for the PES calculated using WOM. Typically, the fastest electrons are missing because they arrive earlier at the absorbing boundary. However, electrons, after substantial excursions, may return to the ion due to the oscillatory laser field, and scatter. If the numerical grid is so small that parts of the wavefunction representing such electrons are absorbed, the PES may be spoiled not only at high energies.

In order to calculate the energy-resolved PES the window operator
\begin{equation}
  \label{eq:window-op}
  \hat W_{\gamma n}(\epsilon)=\frac{\gamma^{2^n}}{{(\hat H_0-\epsilon)}^{2^n}+\gamma^{2^n}}
\end{equation}
is applied to the final state $\ket{\Psi}=\ket{\Psi(t_\mathrm{f})}$ (after the interaction with the external field). $\hat H_0$ is the field-free Hamiltonian (i.e., \eqref{eq:hamiltonian-3D} with $\vecA \equiv \mathbf{0}$), $\gamma$ is the window width, $n$ the window ``order'' (the higher $n$, the more rectangular the window; $n=3$ is used in \qprop{}). The WOM provides an approximation for the absolute squares of the expansion coefficients in energy eigenstates $|c_{\epsilon}|^2=|\braket{\epsilon}{\Psi}|^2$ as
\begin{equation}
  \label{eq:window-limit}
  \lim_{\gamma\rightarrow 0} \frac{1}{N_{\gamma n}} \expect{\Psi}{\hat W_{\gamma n }^2(\epsilon)}=|c_{\epsilon}|^2, \qquad N_{\gamma n}=\int \diff\epsilon \left(\frac{\gamma^{2^n}}{\epsilon^{2^n}+\gamma^{2^n}}\right)^2.
\end{equation}
Application of the window operator to the final wavefunction, 
\begin{equation}
  \label{eq:window-apply}
  \ket{\chi_{\gamma n}(\epsilon)} = \hat W_{\gamma n}(\epsilon) \ket{\Psi}, \qquad \braket {\vecr}{\chi_{\gamma n}(\epsilon)} = \frac{1}{r}\sum_{\ell m} R_{\ell m}^{(\gamma n)}(\epsilon,r) Y_{\ell m}(\Omega),
\end{equation}
allows to calculate the energy-differential ionization probability as (dropping $\gamma n$)
\begin{eqnarray}
  \label{eq:window-prob}
  \abl{P(\epsilon)}{\epsilon} &=& \braket{\chi(\epsilon)}{\chi(\epsilon)}
  = \iint\! \diff r\,\diff\Omega\, \Big\vert\sum_{\ell m} R_{\ell m}(\epsilon,r) Y_{\ell m}(\Omega) \Big\vert^2 
  = \sum_{\ell m} \int \diff r \big\vert R_{\ell m}(\epsilon,r) \big\vert^2 \nonumber \\
  &=:& \sum_{\ell m} \big\vert a_{\text{winop},\ell m}(\epsilon) \big\vert^2.
\end{eqnarray}
Omitting the integration over the angles $\theta$, $\varphi$ in $\Omega$ allows to approximate energy-angle-differential spectra,
\begin{equation}
  \label{eq:window-prob-energ-angle}
  \frac{\diff P(\epsilon,\Omega)}{\diff\Omega\,\diff\epsilon}  
  = \int \diff r\,\Big\vert\sum_{\ell m} R_{\ell m}(\epsilon,r) Y_{\ell m}(\Omega) \Big\vert^2. \end{equation}
This is an approximation because immediately after the laser pulse the solid angle $\Omega$ in position space is not yet equal to the emission solid angle determined by the asymptotic electron  momentum. Hence, in practice it is advisable to field-free post-propagate a while after the laser pulse until the energy-angle-differential spectrum is converged (the lower the energy, the longer it takes for convergence). Note that because of $[\hat W_{\gamma n}(\epsilon),\hat H_0]=0$ the angle-integrated spectrum  \eqref{eq:window-prob} is converged immediately after the pulse. 

As mentioned above, long pulse durations and high electron momenta render WOM very costly
because all of the rapidly spreading wavefunction has to be retained on the numerical grid. 

\subsection{Photoelectron spectra with t-SURFF}\label{sec:tSURFF}
In order to facilitate the calculation of momentum-resolved spectra on smaller spatial grids the t-SURFF method was proposed~\cite{tsurff}.
For simplicity and pedagogical reasons, let us first consider the one-dimensional TDSE
\begin{equation}
  \label{eq:one-d-sgl}
  \imagi\partial_t \Psi(x,t) = \left( -\frac{1}{2}\partial_x^2-\imagi A(t)\partial_x + V(x) \right) \Psi(x,t) .
\end{equation}
Suppose the  binding potential $V(x)$ can be neglected for distances $\vert x\vert > \XI >0$, and the propagation lasts  long enough, i.e., up to time $t=T$ when the laser is off and 
all probability density representing ionization with a certain final minimum electron momentum arrived at $\pm\XI$.
Then the probability amplitude for ionization can be approximated by
\begin{equation}
  \label{eq:one-d-ion-amplitude}
  a_{\text{I}}(k) =
  \langle k(T) \vert \Theta(\vert x \vert - \XI) \vert \Psi(T) \rangle 
  =\int \!\diff x\; \Theta(\vert x \vert - \XI) \psi^*_k(x,T) \Psi(x, T) 
\end{equation}
with plane-wave final-momentum states  $\ket{k(T)}$, i.e., in position space  $\psi_k(x,T)=\braket{x}{k(T)}$.
We proceed by apparent complication, writing
\begin{equation}
  \label{eq:one-d-ion-amplitude-dt}
  a_{\text{I}}(k)
  =\int\limits_{0}^{T} \!\diff t\, \partial_t \langle k(t) | \Theta(\vert x \vert -\XI) | \Psi(t) \rangle 
  + \langle k(0) |\Theta(\vert x \vert -\XI) | \Psi(0) \rangle .
\end{equation}
Let us assume that the laser is on within the time interval $[0,\TP]$, $\TP\leq T$, turning   $\vert k(t) \rangle$ into a Volkov state \cite{volkov35,sergeyreview2014} for the TDSE  \eqref{eq:one-d-sgl} with $V(x)\equiv 0$, that is, the solution  for a free electron in a laser field. In position space and velocity gauge (with $A^2(t)$ transformed away) the Volkov state reads
\begin{equation}
  \label{eq:volkov_oneD}
  \psi_{k}(x,t)={(2\pi)}^{-1/2} \eulere^{- \imagi k^2 t/2 + \imagi k [x-\alpha(t)]}, \qquad \alpha(t) = \int_0^t \diff t'\, A(t') .
\end{equation}
$\alpha(t)$ is the classical excursion of a free electron in the laser field.
Using the TDSE  \eqref{eq:one-d-sgl} in \eqref{eq:one-d-ion-amplitude-dt}, the fact that $V(x)\simeq 0$ for $|x|> \XI$, and $\langle k(0) | \Theta(\vert x \vert -\XI) | \Psi(0) \rangle \simeq 0$ for bound initial states, we obtain an expression with the commutator between the Volkov-Hamiltonian $-\frac{1}{2}\partial_x^2-\imagi A(t)\partial_x$ and the t-SURFF-boundary-defining step function, 
\begin{equation}
  \label{eq:one-d-ion-amplitude-tsurff}
  a_{\text{I, t-SURFF}}(k)
  =\imagi \int\limits_{0}^{T} \!\diff t\,  \langle k(t) |\left[-\frac{1}{2}\partial_x^2-\imagi A(t)\partial_x\, ,\, \Theta(\vert x \vert -\XI)\right] | \Psi(t) \rangle .
\end{equation}
As $\int \!\diff x\, \partial_x \theta(\vert x\vert-X)=\int \!\diff x\,  [\delta(x-X)-\delta(x+X)]$ and 
$\int \!\diff x\, f(x)\partial_x \delta(x-X)=-\int \!\diff x\, \delta(x-X)\partial_x f(x)$, we find---with the Volkov states inserted---
\begin{multline}
  \label{eq:one-d-ion-amplitude-tsurff-2}
  a_{\text{I, t-SURFF}}(k) 
=  \frac{1}{\sqrt{2\pi}} \left[   \int_0^T \diff t\, \eulere^{\imagi t k^2/2}   \ \eulere^{-\imagi k[x-\alpha(t)]}\left\{\halb k +  A(t) - \frac{\imagi}{2}\pabl{}{x} \right\}  \Psi(x,t) \right]_{x=-\XI}^{\XI} ,
\end{multline}
and the momentum-resolved spectrum follows as $ \diff P(k)/\diff k=\vert a_{\text{I, t-SURFF}}(k) \vert^2$.

In order to avoid finite-$T$-dependent artifacts  half a Hanning window 
\begin{equation}
  \label{eq:hanning_window}
H(t)=
  \begin{cases}
   1 & \text{if } t < T/2 \\
   [1-\cos(2\pi t/T)]/2       & \text{if } t \geq T/2 
  \end{cases}  
\end{equation}
may be multiplied to the integrands.

$\XI$ should be big enough so that $|V(x)|$ is sufficiently small for $|x|>\XI$. On the other hand, t-SURFF only captures electrons represented by the parts of the wavefunction that leave the region $|x|<\XI$ within the time interval $[0,T]$. In practice, a compromise has to be found, and the convergence of the spectra in the momentum range of interest should be checked by varying $\XI$ and $T$.

\subsubsection{Angle-momentum-resolved spectra with \qprop{}}\label{sec:tSURFF-threed}
In the one-dimensional case, t-SURFF amounts to analyzing the flux through a surface consisting of only two points $\pm\XI$. In three-dimensional problems the role of $\XI$ may be taken by a radius $\RI$, and the binding potential $V(\vecr)$ is then assumed to be negligible for $|\vecr| = r >\RI$. The analogue of  \eqref{eq:one-d-ion-amplitude-tsurff-2} then involves integrals $\int\diff\Omega$ over the surface of the sphere of radius $\RI$.  Instead of   \eqref{eq:one-d-ion-amplitude-tsurff} we now have
\begin{multline}
  \label{eq:three-d-ion-amplitude-tsurff}
  a_{\text{I, t-SURFF}}(\veck)
  = \imagi \int\limits_{0}^{T} \diff t\, \langle \veck(t) | \left[-\frac{1}{2} \grad^2 -\imagi \vecA(t)\cdot\grad ,\  \Theta( r  -\RI)\right] | \Psi(t) \rangle \\
= \RI^2 \int\limits_0^T \diff t\, \int \diff\Omega \Bigg\lbrace \psi_{\veck}^*(\vecr, t) [ A_x(t) \sin\theta\cos\varphi 
    + A_y(t) \sin\theta\sin\varphi \\
    + A_z(t) \cos\theta ] \Psi(\vecr, t) 
-\frac{\imagi}{2} [ \psi_{\veck}^*(\vecr, t) \partial_r \Psi(\vecr, t) - \Psi(\vecr, t) \partial_r \psi_{\veck}^*(\vecr, t) ] \Bigg\rbrace_{r=\RI},
\end{multline}
with Volkov waves
\begin{equation}
  \label{eq:threeD-volkov-wave1}
  \psi_{\veck}(\vecr, t) = (2\pi)^{-3/2} \eulere^{-\imagi k^2t/2 + \imagi \veck \cdot [\vecr -\vecalpha(t)]}, \qquad  \vecalpha(t)=\int_0^t \diff t'\, \vecA(t').
\end{equation}
Apart from the transformation~\eqref{eq:app-trafo} (and a different definition of the sign of the electron charge in atomic units) this is the same result for the probability amplitudes $a(\veck)$ as in Ref.~\cite{tsurff}. 
In Qprop, we need to  calculate the surface integral
in the probability amplitude~\eqref{eq:three-d-ion-amplitude-tsurff} for the case where the time-dependent wave function is available
as an expansion in spherical harmonics~\eqref{eq:wave-fun-expansion_elliptpol}.
To this end we use an expansion of the Volkov waves

\begin{equation}
  \label{eq:threeD-volkov-wave2}
  \psi_{\veck}(\vecr, t)  = \sqrt{\frac{2}{\pi}}\ \eulere^{-\imagi k^2t/2 - \imagi \veck \cdot \vecalpha(t)} \sum_{\ell m} \imagi^{\ell} j_{\ell}(k r) Y^*_{\ell m}(\Omega_k) Y_{\ell m}(\Omega).
\end{equation}
Here, $\Omega$ is the solid angle with respect to $\vecr$, the solid angle $\Omega_k$ is with respect to to $\veck$, and  $j_{\ell}(k r)$ are the spherical Bessel functions.
Inserting \eqref{eq:threeD-volkov-wave2} and the spherical harmonics expansion  \eqref{eq:wave-fun-expansion_elliptpol} into \eqref{eq:three-d-ion-amplitude-tsurff}
yields
\begin{multline}
  \label{eq:three-d-ion-amplitude-tsurff-ellm-1a}
  a_{\text{I, t-SURFF}}(\veck) \\
  = \sqrt{\frac{2}{\pi}} \RI^2 \int\limits_0^T \!\diff t\, \eulere^{\imagi k^2t/2 + \imagi \veck \cdot \vecalpha(t)} \int \diff \Omega 
  \sum_{\ell m,\ell_1 m_1} {(-\imagi)}^{\ell_1} Y_{\ell_1 m_1}(\Omega_k) Y_{\ell_1 m_1}^*(\Omega) Y_{\ell m}(\Omega) \\
  \times \Bigg\lbrace j_{\ell_1}(k r) \sqrt{\frac{2\pi}{3}} 
  \left[  \tilde{A}(t)Y_{1,-1}(\Omega)  - \tilde{A}^*(t)Y_{1,1}(\Omega) +  \sqrt{2} A_z(t) Y_{1,0}(\Omega)  \right]  \frac{1}{r}  \phi_{\ell m}(r,t) \\ 
  - \frac{\imagi}{2} j_{\ell_1}(k r) \partial_r \left[ \frac{1}{r} \phi_{\ell m}(r,t)  \right] 
  + \frac{\imagi}{2 r} \phi_{\ell m}(r,t) \partial_r j_{\ell_1}(k r)\Bigg\rbrace_{r=\RI} .
\end{multline}
Here, $\tilde A(t) = A_x(t)+\imagi A_y(t)$.
The solid-angle integrals over three spherical harmonics with argument $\Omega$ can be evaluated. One obtains 
\begin{equation}
  \label{eq:expansion1}
  a_{\text{I, t-SURFF}}(\veck) = \sum_{\ell m} a_{\text{I, t-SURFF}, \ell m}(\veck) Y_{\ell m}(\Omega_k) 
\end{equation}
with
\begin{multline}
  \label{eq:three-d-ion-amplitude-tsurff-ellm-4}
a_{\text{I, t-SURFF}, \ell m}(\veck)  \\
 = \frac{\RI (-\imagi)^{\ell+1}}{(2\pi)^{1/2}}      \int_0^T \diff t\, \eulere^{\imagi t k^2/2+\imagi \veck\cdot\vecalpha(t)} \Biggl\{   j_{\ell}(k\RI)  \left[ \left.\partial_r \phi_{\ell m}(r,t) \right|_\RI  -  \frac{(\ell+1)\phi_{\ell m}(\RI,t)}{\RI} \right] \\
\qquad\qquad\qquad\qquad + k\,  \phi_{\ell m}(\RI,t)  j_{\ell+1}(k\RI) \\
  +\imagi \sqrt{2}\, j_{\ell}(k\RI) \left[ \tilde A(t) \Biggl(  b_{\ell,-m}\phi_{\ell-1,m+1}(\RI,t) -   d_{\ell m} \phi_{\ell+1,m+1}(\RI,t)\Biggr) \right.\\
\qquad\qquad\qquad\qquad-\tilde A^*(t) \Biggl( b_{\ell m}\phi_{\ell-1, m-1}(\RI,t) -  d_{\ell, -m}\phi_{\ell+1, m-1}(\RI,t) \Biggr) \\
\left. + \sqrt{2} A_z(t) \Biggl(c_{\ell-1,m}   \phi_{\ell-1,m}(\RI,t)  + c_{\ell m}  \phi_{\ell+1,m}(\RI,t)\Biggr) \right]
\Biggr\} ,
\end{multline}
where the  the recursion relation
\begin{equation}
  \label{eq:bessel-recurse}
  \abl{}{r} j_{\ell}(kr) = -k j_{\ell+1}(kr) + \frac{\ell}{r} j_{\ell}(kr)
\end{equation}
was used,
and
\begin{align}
c_{\ell m} &= \sqrt{\frac{(\ell+1)^2-m^2}{(2\ell+1)(2\ell+3)}},\qquad b_{\ell m} =\sqrt{\frac{(\ell+m-1)(\ell+m)}{2(2\ell-1)(2\ell+1)}},\\
 d_{\ell m} &=\sqrt{\frac{(\ell +m+1)(\ell +m  +2)(\ell+1)}{(2\ell+2)(2\ell+3)(2\ell+1)}}.
\end{align}
The time integrals at the surface
\begin{align}
  \label{eq:int-dt-0}
  I_{i, \ell m}(\veck) &= \int\limits_0^T \!\diff t\, \eulere^{\imagi t k^2/2+\imagi \veck\cdot\vecalpha(t)} F_i(t) \phi_{\ell m}(\RI,t), \qquad i=0,1,2,3\\
  I_{4, \ell m}(\veck) &= \int\limits_0^T \!\diff t\, \eulere^{\imagi t k^2/2+\imagi \veck\cdot\vecalpha(t)} \partial_r \phi_{\ell m}(r,t)|_{r=\RI} \label{eq:int-dt-0other}
\end{align}
with
\begin{equation} F_0(t)= \tilde{A}^*(t), \quad  F_1(t)= \tilde{A}(t), \quad  F_2(t)= {A}_z(t), \quad F_3=1 \end{equation}
are needed to calculate the t-SURFF spectrum,
\begin{multline}
  \label{eq:three-d-ion-amplitude-tsurff-ellm-5}
a_{\text{I, t-SURFF}, \ell m}(\veck)  = \frac{\RI (-\imagi)^{\ell+1}}{(2\pi)^{1/2}}      \Biggl\{   j_{\ell}(k\RI)  \left[
 I_{4, \ell m}(\veck) -  \frac{\ell+1}{\RI} I_{3, \ell m}(\veck) \right] + k\,  j_{\ell+1}(k\RI)  I_{3, \ell m}(\veck)  \\
  +\imagi \sqrt{2}\, j_{\ell}(k\RI) \left[- ( b_{\ell m} I_{0, \ell-1, m-1}(\veck)
 -  d_{\ell,-m} I_{0, \ell+1, m-1}(\veck)
 ) \right.\\
\qquad\qquad\qquad\qquad +  (  b_{\ell,- m}  I_{1, \ell-1, m+1}(\veck)
 -  d_{\ell m}  I_{1, \ell+1, m+1}(\veck)
)  \\
 \left. + \sqrt{2}  (  c_{\ell-1,m}  I_{2, \ell-1, m}(\veck)  + c_{\ell m}  I_{2, \ell+1, m}(\veck) ) \right]
\Biggr\} . 
\end{multline}

In the expansion of the probability amplitude~\eqref{eq:expansion1}, $a_{\text{I, t-SURFF}, \ell m}(\veck)$ still depends on $\Omega_k$. 
 In a ``complete'' expansion in spherical harmonics one would expect no angular dependence in the coefficients, i.e.,
\begin{equation}
  \label{eq:expansion2}
  a_{\text{I, t-SURFF}}(\veck) = \sum_{\ell m} \abar_{\text{I, t-SURFF}, \ell m}(k) Y_{\ell m}(\Omega_k) .
\end{equation}
This can be achieved by expanding
\begin{equation} 
\eulere^{\imagi \veck\cdot\vecalpha(t)} =  4\pi \sum_{\ell m} \imagi^{\ell} j_{\ell}[k\alpha(t)] Y_{\ell m}^*(\Omega_{\alpha(t)}) Y_{\ell m}(\Omega_{k}) 
\end{equation}
as well. Another solid angle $\Omega_{\alpha(t)}$, with respect to the excursion vector $\vecalpha(t)$, appears, and 
\begin{multline} \label{eq:abar1}
 \abar_{\text{I, t-SURFF}, \ell m}(k) =   \RI  \sum_{\ell_1 m_1\ell_2} 
 \sqrt{\frac{2(2\ell_1+1)(2\ell_2+1)}{(2l+1)}} C_{\ell_10\ell_20}^{\ell 0}  C_{\ell_1 m_1\ell_2,m-m_1}^{\ell m}\\
\times  (-\imagi)^{\ell_1-\ell_2+1}   \int_0^T \diff t\, \eulere^{\imagi t k^2/2}\ j_{\ell_2}[k\alpha(t)] Y_{\ell_2,m-m_1}^*(\Omega_{\alpha(t)}) \\
\quad\times 
 \Biggl\{    j_{\ell_1}(k\RI)  \biggl[ \partial_r \phi_{\ell_1 m_1}(r,t) \biggr|_\RI  -\frac{(\ell_1+1)\phi_{\ell_1 m_1}(\RI,t)}{\RI} \biggr] + k\,  \phi_{\ell_1 m_1}(\RI,t)  j_{\ell_1+1}(k\RI)   \\
+ \imagi \sqrt{2}\, j_{\ell_1}(k\RI) \biggl[ \tilde A(t) \bigl(  b_{\ell_1,-m_1}
\phi_{\ell_1-1,m_1+1}(\RI,t) -  d_{\ell_1 m_1} \phi_{\ell_1+1,m_1+1}(\RI,t)\bigr) \\
 -\tilde A^*(t) \bigl( b_{\ell_1 m_1}
\phi_{\ell_1-1,m_1-1}(\RI,t)  - d_{\ell_1,-m_1} \phi_{\ell_1+1,m_1-1}(\RI,t) \bigr)\\
 +\sqrt{2} A_z(t) \bigl(c_{\ell_1-1,m_1}   \phi_{\ell_1-1,m_1}(\RI,t)  + c_{\ell_1 m_1}  \phi_{\ell_1+1,m_1}(\RI,t)\bigr)
\biggr]\Biggr\},
\end{multline}
where $C_{\ell_1 m_1\ell_2 m_2}^{\ell m}$ are Clebsch-Gordan coefficients \cite{varshal}, is obtained.
The relevant time integrals now read
\begin{align}
  \label{eq:int-dt-new-0}
  \Ibar_{i,\ell_1 m_1,\ell_2 m_2}(k) &= \int_0^T \!\diff t\,    \eulere^{\imagi k^2t/2}  j_{\ell_2}[k \alpha(t)] Y^*_{\ell_2 m_2}(\Omega_{\alpha(t)}) F_i(t) \phi_{\ell_1 m_1}(\RI,t), \quad i=0,1,2,3,\\
  \Ibar_{4,\ell_1 m_1,\ell_2 m_2}(k) &= \int_0^T \!\diff t\,    \eulere^{\imagi k^2t/2}  j_{\ell_2}[k \alpha(t)] Y^*_{\ell_2 m_2}(\Omega_{\alpha(t)}) \partial_r \phi_{\ell_1 m_1}(r,t)|_{\RI}, \label{eq:int-dt-new-0other}
\end{align}
in terms of which
\begin{multline}\label{eq:three-d-ion-amplitude-tsurff-ellm-6}
 \abar_{\text{I, t-SURFF}, \ell m}(k) =   \RI  \sum_{\ell_1 m_1\ell_2} 
 \sqrt{\frac{2(2\ell_1+1)(2\ell_2+1)}{(2l+1)}} C_{\ell_10\ell_20}^{\ell 0}  C_{\ell_1 m_1\ell_2,m-m_1}^{\ell m} (-\imagi)^{\ell_1-\ell_2+1}   \\
\quad\times
 \Biggl\{    j_{\ell_1}(k\RI)  \left[  \Ibar_{4,\ell_1 m_1,\ell_2,m-m_1}(k)  -  \frac{\ell_1+1}{\RI} \Ibar_{3,\ell_1 m_1,\ell_2,m-m_1}(k) \right]  \\
+ k\,  \Ibar_{3,\ell_1 m_1,\ell_2,m-m_1}(k)  j_{\ell_1+1}(k\RI)   \\
\qquad  + \imagi \sqrt{2}\, j_{\ell_1}(k\RI) \left[
   b_{\ell_1,-m_1} \Ibar_{1,\ell_1-1,m_1+1,\ell_2,m-m_1}(k) -  d_{\ell_1 m_1}
 \Ibar_{1,\ell_1+1,m_1+1,\ell_2,m-m_1}(k) \right.\\
 \qquad\qquad\qquad\qquad - b_{\ell_1 m_1}
 \Ibar_{0,\ell_1-1,m_1-1,\ell_2,m-m_1}(k) +   
d_{\ell_1,-m_1} \Ibar_{0,\ell_1+1,m_1-1,\ell_2,m-m_1}(k) 
 \\
\left. + \sqrt{2}  \bigl(  c_{\ell_1-1,m_1}  \Ibar_{2,\ell_1-1,m_1,\ell_2,m-m_1}(k) + c_{\ell_1 m_1} \Ibar_{2,\ell_1+1,m_1,\ell_2,m-m_1}(k) \bigr) \right]\Biggr\}
\end{multline}
results.

Both methods for calculating the ionization probability amplitude, i.e., via \eqref{eq:expansion1} with  \eqref{eq:three-d-ion-amplitude-tsurff-ellm-5} and \eqref{eq:expansion2} with  \eqref{eq:three-d-ion-amplitude-tsurff-ellm-6}, 
are implemented in \qprop~2.0.
If $N_{\theta_k}$, $N_{\varphi_k}$ are the number of respective angles, and $N_{\ell}=L_{\max}$, $N_{m}$ the number of $\ell$ and $m$ quantum numbers considered, the ratio $N_{\theta_k}N_{\varphi_k}/(N_{\ell} N_{m})$ of the number of time integrals that need to be calculated may be used to estimate which of the two methods is computationally cheaper.

The energy-differential ionization probability $ \diff P_{\text{I, t-SURFF}}(\epsilon)/\diff \epsilon$ with $\epsilon=k^2/2$ can be calculated (using
$
\diff^3 k  = k^2 \diff k \,\diff\Omega_k =  \sqrt{2\epsilon} \,\diff \epsilon\,\diff\Omega_k$)
as
\begin{multline}
  \label{eq:tsurff-total}
 \abl{ P_{\text{I, t-SURFF}}(\epsilon)}{\epsilon} = \sqrt{2\epsilon}\int \diff\Omega_k\, \Big\vert \sum_{\ell m} \abar_{\text{I, t-SURFF},\ell m}(k) Y_{\ell m}(\Omega_k) \Big\vert^2 \Bigg\vert_{k=\sqrt{2\epsilon}} \\
  =\sqrt{2\epsilon} \sum_{\ell m} \big\vert \abar_{\text{I, t-SURFF},\ell m}(k) \big\vert^2  \Bigg\vert_{k=\sqrt{2\epsilon}}  =\sqrt{2\epsilon}  |\abar_{\text{I, t-SURFF}}(k)|^2 \Bigg\vert_{k=\sqrt{2\epsilon}}.
\end{multline}
The last expression enables a direct comparison of the partial spectra $|\abar_{\text{I, t-SURFF},\ell m}(k)|^2$ 
with the WOM result~\eqref{eq:window-prob}.

The two propagation modes implemented in \qprop\ cover linear polarization, $A_z\neq0$, $\tilde A=\tilde A^*\equiv 0$ (mode {\tt 34}) and elliptical polarization in the $xy$-plane, $\tilde A\neq 0$, $\tilde A^*\neq 0$, $A_z\equiv 0$   (mode {\tt 44}). The corresponding t-SURFF spectral amplitudes follow from the more general expressions \eqref{eq:three-d-ion-amplitude-tsurff-ellm-5}, \eqref{eq:three-d-ion-amplitude-tsurff-ellm-6}.

\section{News in \qprop~2.0}\label{sec:lib}
The structure of \qprop{}, propagation modes, output, WOM analysis etc.\ are described in the original \qprop{} article~\cite{bauer2006qprop}. 
In a typical TDSE-solving problem, an imaginary-time propagation to find the initial state precedes a real-time propagation of the wavefunction. After the real-time propagation, the final wavefunction may be analyzed. Since the earliest versions of \qprop{} WOM was implemented to calculate photoelectron spectra.  Now, in \qprop~2.0, there is an alternative to the last, WOM step, which is t-SURFF. However, while WOM requires the final wavefunction and the binding potential only,  t-SURFF  needs data stored during the real-time propagation as well, and the real-time propagation depends on where the t-SURFF boundary $\RI$ is located. In the example of Section   \ref{sec:first-example} WOM and t-SURFF spectra will be calculated and compared.  Before, we briefly mention other important changes in \qprop~2.0.

External potentials are still collected in the class \verb!hamop!.  Up to now this class could only handle functions (cf.\ Table 2 in~\cite{bauer2006qprop}). In \qprop~2.0 it is able to digest any object that can be converted to \verb!std::function!. In the examples in Section \ref{sec:examples} this is exploited by using functors instead of functions.

The class \verb!parameterListe! is provided for parsing simple parameter files.
These text files contain entries of the form
\verb! name type value!.
Lines starting with the character {\tt \#} are ignored and can be used for comments.
In order to read parameters from a file functions for reading the types \verb!string!, \verb!long! and \verb!double! are implemented.
The source code of the test cases  in Section \ref{sec:examples} provide plenty of examples for the use of parameter files. 

The t-SURFF method for calculating PES is implemented in the classes \verb!tsurffSpectrum! and \verb!tsurffSaveWF!.
The class \verb!tsurffSaveWF! is responsible  for saving the radial wavefunctions at the t-SURFF boundary $\phi_{\ell m}(\RI,t)$ and their spatial derivative $\partial_r\phi_{\ell m}(r,t)\vert_{r=\RI}$  (fourth order finite difference approximation) to files with the ending {\tt .raw}.

The remaining steps, i.e.,
performing the time integrals \eqref{eq:int-dt-0}, \eqref{eq:int-dt-0other} or  \eqref{eq:int-dt-new-0},  \eqref{eq:int-dt-new-0other} (smoothed by the Hanning window~\eqref{eq:hanning_window}) and calculating the partial spectra~\eqref{eq:expansion1} or~\eqref{eq:expansion2}
are implemented in the class \verb!tsurffSpectrum!.
The relevant member functions are \verb!time_integration()! and \verb!polar_spectrum()!, respectively.

In \qprop\ 2.0, the class {\tt vecpot}---to be defined in {\tt potentials.hh}---is used to initialize the vector potential components for the real-time propagation. The examples below illustrate this.

Depending on the parameter \verb!expansion-method! in \verb!tsurff.param!,  equation~\eqref{eq:expansion1} (\verb!expansion-method!=1) 
or~\eqref{eq:expansion2} (\verb!expansion-method!=2) is employed to calculate the probability amplitudes.
If expression~\eqref{eq:expansion2} is used \verb!print_partial_amplitudes()!
may be called to write the partial amplitudes $\sqrt{2\epsilon}|\abar_{\text{I, t-SURFF},\ell m}(\sqrt{2\epsilon})|^2$ to a file.

As in the previous versions of \qprop\ there are  two possible field set-ups: linear polarization along the $z$-direction (propagation mode {\tt 34}) and any polarization in the $xy$-plane (propagation mode {\tt 44}). 
They are selected by {\tt qprop-dim long 34} or {\tt qprop-dim long 44} in the parameter file {\tt initial.param}, respectively.

Angle-resolved spectra whose range and resolution are defined in the parameter file {\tt tsurff.param} are written to text files named \verb!tsurff-polar!$i_{\text{p}}$\verb!.dat!. 
 Here, $i_{\text{p}}$ is the number of the process which produced the result. By default MPI parallelization is disabled and there is only a single file with $i_{\text{p}}=0$. However, the examples in  Section \ref{sec:examples} can be also processed using a parallel t-SURFF analysis.
For the case of polarization in the $xy$-plane each data row contains the energy value $k^2/2$, absolute value of momentum $k$, angle $\theta_k$, angle $\varphi_k$ and amplitude $\vert a(\veck)\vert^2 k$. In the case of linear polarization the  azimuthal angle $\varphi_k$ is not relevant due to the azimuthal symmetry about the $z$ axis, and thus omitted.

The partial spectra files named \verb!tsurff-partial!$i_{\text{p}}$\verb!.dat! (generated if \verb!expansion-method! is set to {\tt 2}) contain the column entries 
energy $k^2/2$, momentum $k$, partial probabilities $\vert \abar_{0,0}(k) \vert^2 k,  \dots, \vert \abar_{L_{\max}-1,L_{\max}-1}(k) \vert^2 k$, and their sum $\vert \abar(k) \vert^2 k$.
The ordering  of the entries $\vert \abar_{\ell m}(k)\vert^2$ for propagation mode {\tt 44} is indicated in Table~\ref{tab:ell-m-order}.
\begin{table}[h]
  \caption{\label{tab:ell-m-order}Mapping of $\ell$ and $m$ to a single index  $(\ell+1)\ell+m$.}
  \begin{tabular}[]{c|ccccccc}
    & $\cdots$ & $m=-2$ & $m=-1$ & $m=0$ & $m=1$ & $m=2$ & $\cdots$  \\
    \hline
    $\ell=0$ & & & & 0 & & & \\
    $\ell=1$ & & & 1 & 2 & 3 & & \\
    $\ell=2$ & & 4 & 5 & 6 & 7 & 8 & \\
    \vdots &  $\udots$ & & & \vdots& & & $\ddots$
  \end{tabular}
\end{table}

In the case of linear polarization along the $z$ axis (propagation mode {\tt 34}) the magnetic-quantum-number $m$ is fixed, and each row has the column entries
$k^2/2$, $k$, $\vert \abar_{0}(k) \vert^2 k \dots \vert \abar_{L_{\max}-1}(k) \vert^2 k$, $\vert \abar(k) \vert^2 k$.
Note that all partial spectra are multiplied by $k=\sqrt{2\epsilon}$ (cf.\ eq.~\eqref{eq:tsurff-total}).

\begin{figure}[h]
  \includegraphics[width=0.76\textwidth]{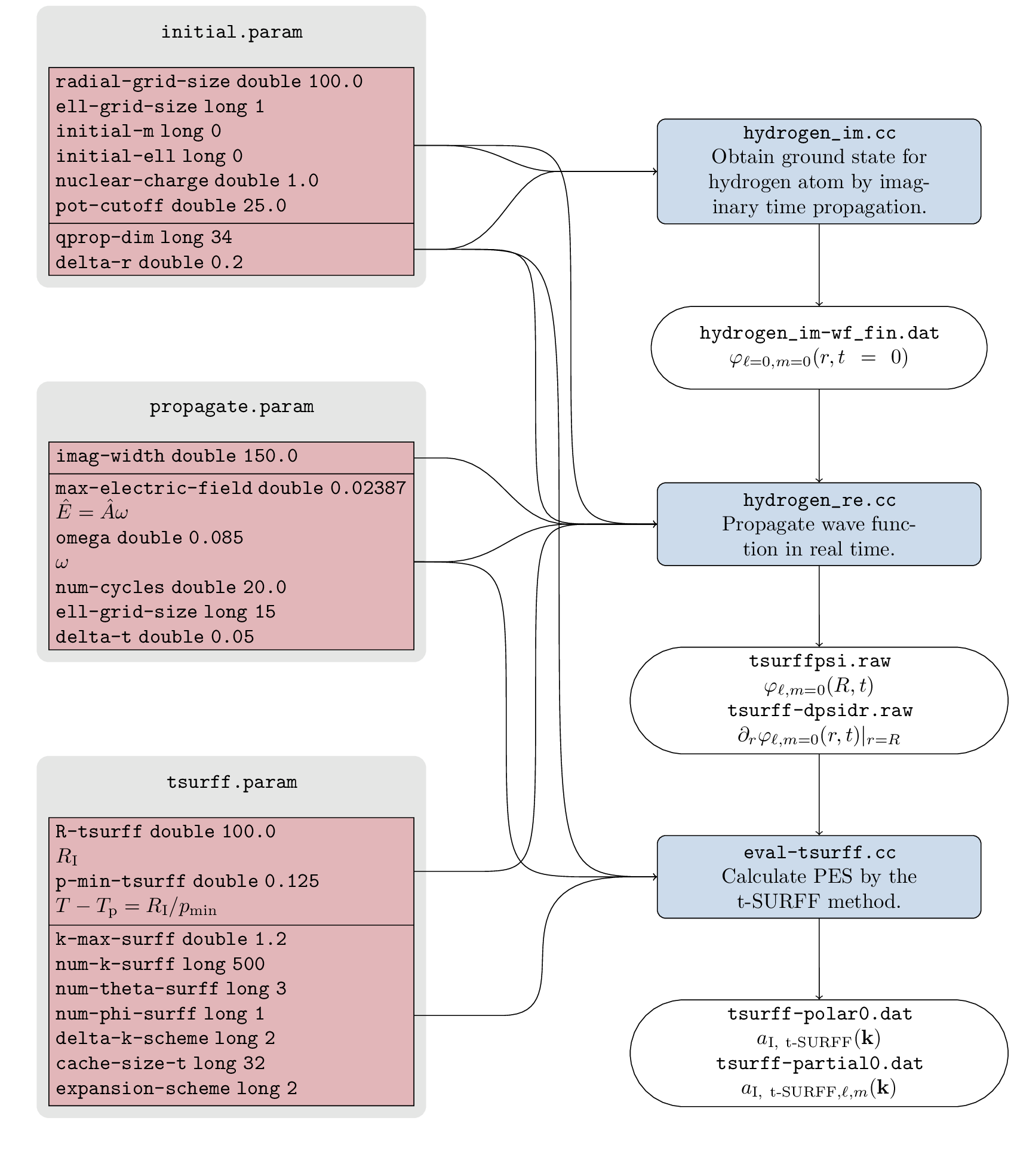}
  \caption{(Color online) The flow chart shows which input parameters (left) are used by which programs (right, light blue). The programs  generate output, some of which is read by another program (right, white). This particular example is for the case discussed in Section~\ref{sec:first-example}, generating the data shown in Figs.~\ref{fig:winop-vs-tsurff} and~\ref{fig:ati-partial}.}
  \label{fig:flow-chart}
\end{figure}

The range and the resolution of the spectra are determined by the following parameters in {\tt tsurff.param}:
\verb!k-max-surff! determines the maximum absolute value of momentum, \verb!num-k-surff! the number of $k$ values for which probabilities are calculated. Setting the parameter \verb!delta-k-scheme! to {\tt 1} samples equidistantly in $k$, \verb!2! equidistantly in energy $\epsilon=k^2/2$. \verb!num-theta-surff! and \verb!num-phi-surff! define the numbers $N_{\theta_k}$, $N_{\varphi_k}$ of angles $\theta_k$ and $\varphi_k$.
The values for the angles are distributed equidistantly in the intervals $\theta_k\in [0,\pi]$ and $\varphi_k\in [0,2\pi)$ (note that if $N_{\theta_k}<3$ it is increased to $3$, and if $N_{\theta_k}$ is chosen even it is increased by one; in that way $\theta_k=0,\pi/2,\pi$ are always covered).

The calculation of a spectrum may be easily parallelized by assigning to each process a part of the $k$ interval. Open MPI~\cite{openmpi} is used in the current implementation.

The GNU Scientific Library (GSL)~\cite{gsl} is used for the evaluation of spherical harmonics, Bessel functions, and Wigner $3j$ symbols. The latter are related to the Clebsch-Gordan coefficients appearing in  \eqref{eq:abar1} and  \eqref{eq:three-d-ion-amplitude-tsurff-ellm-6} \cite{varshal}.

\begin{table}[h]

  \caption{\label{tab:changes} Comparison of features in \qprop{} and \qprop~2.0}
  \begin{tabular}[]{lll}
    & \qprop{} & \qprop~2.0  \\
    \hline
    PES methods & WOM & t-SURFF and WOM \\
    TDDFT capabilities & yes &  not yet \\
length gauge  &  yes  &  no \\
parallel processing & no & yes (PES with t-SURFF) \\
    representation of potentials & plain functions & std::function \\
    parsing parameter files & xml-like format & name-type-value tuples in text file
  \end{tabular}

\end{table}

\section{Examples}\label{sec:examples}
Four examples for \qprop~2.0 with t-SURFF are provided in the sub-directories \verb!ati-tsurff!, \verb!ati-winop!, \verb!large-clubs!, \verb!attoclock!, and \verb!pow-8-sine!.
Instructions on how to build and run the sample programs and how to plot the results are detailed in the \verb!readme.txt! files provided in these directories.

Simulation parameters are read from the text files \verb!initial.param!, \verb!propagate.param! and \verb!tsurff.param! by the programs for imaginary-time propagation, real-time propagation, and the calculation of PES.
The flow chart in Fig.~\ref{fig:flow-chart} visualizes this for the first example in Section~\ref{sec:first-example}: the parameter files (left) are read by the programs (right) as indicated by lines. Note that both \verb!hydrogen_re.cc! and \verb!eval-tsurff.cc! use parameters from all three parameter files.

Some of the output by one program is read by another, e.g., the ground state wavefunction after imaginary-time propagation in \verb!hydrogen_im-wf_fin.dat! or the wavefunction on the t-SURFF boundary during real time in \verb!tsurffpsi.raw!. The PES data are finally in the files \verb!tsurff-polar0.dat! and \verb!tsurff-partial0.dat!, to be processed by some plot program. In the examples directories gnuplot scripts are provided.

The binding potential $V_{\text{bind}}(r)$, vector potential $\vecA(t)$, excursion $\vecalpha(t)$, and the imaginary potential $-\imagi V_{\text{Im}}(r)$ for absorbing outgoing electron flux (that already passed the t-SURFF boundary)
are defined in the header file \verb!potentials.hh!.
In all examples the imaginary potential is chosen
\begin{equation}
  \label{eq:imag-pot}
  V_{\text{Im}}(r)=
  \begin{cases}
    0 & r < R_{\text{Im}} \\
    V_{\text{Im,max}} {\left(\frac{r-R_{\text{Im}}}{W_{\text{Im}}}\right)}^{16} & r \geq R_{\text{Im}} \\
  \end{cases}
\end{equation}
with $V_{\text{Im,max}}=100$, $R_{\text{Im}}=R_{\text{grid}}-W_{\text{Im}}$, and  the width of the absorbing region $W_{\text{Im}}$ specified via the parameter \verb!imag-width! in {\tt propagate.param}.

An advanced method for the absorption of wavefunctions at grid boundaries with impressively few additional grid points was proposed~\cite{scrinzi-irecs} and could be implemented in a future version of \qprop.


\subsection{Window operator vs. t-SURFF}\label{sec:first-example}

\begin{figure}[t]
  \includegraphics[width=\textwidth]{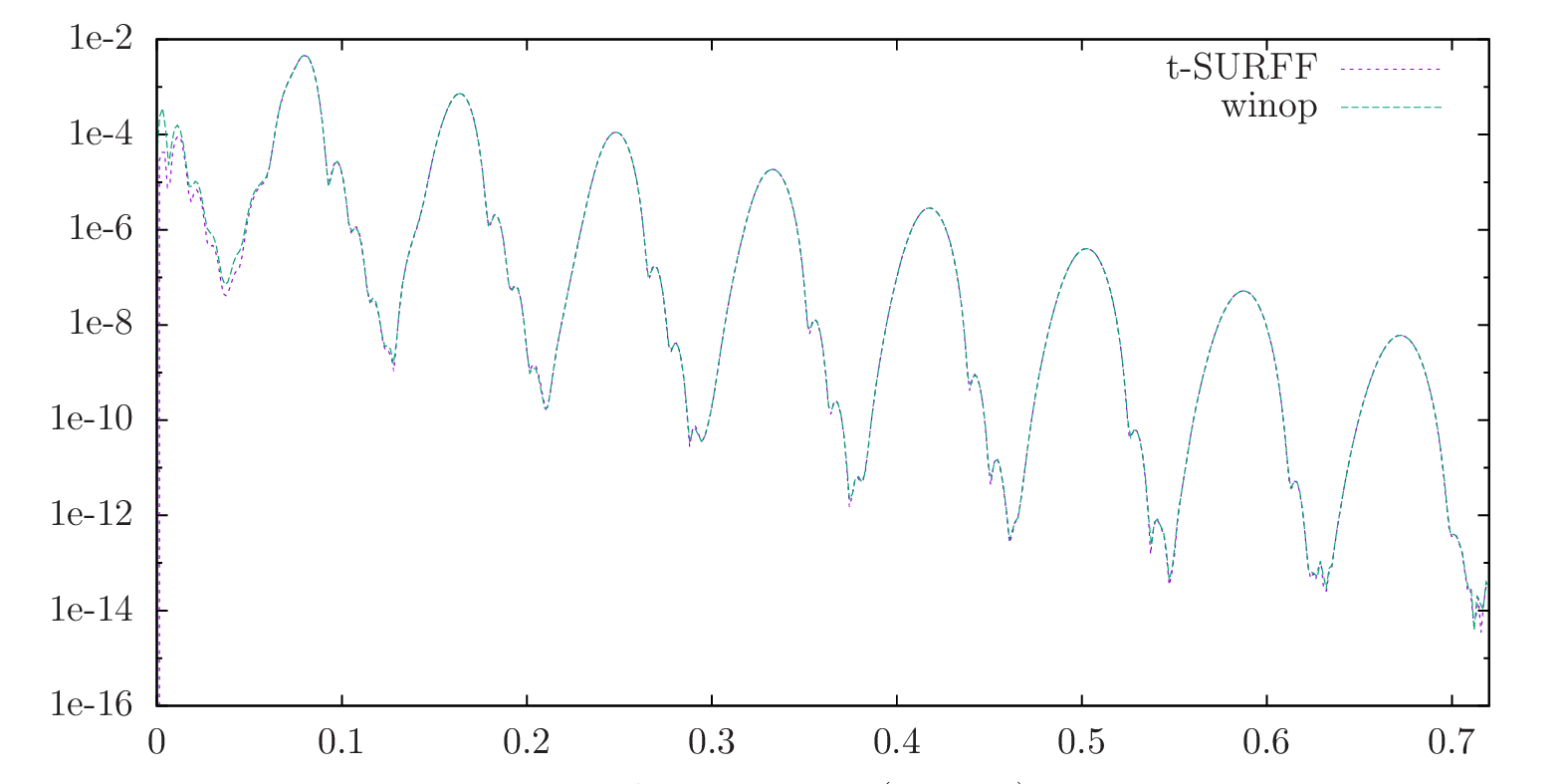}
  \caption{\label{fig:winop-vs-tsurff} (Color online) Energy-resolved total PES for hydrogen (starting from the 1s state) calculated with t-SURFF and WOM. Laser parameters: $\lambda=535$\,nm, $I=2\times 10^{13}\text{ W/cm}^2$, $n_{\text{c}}=20$.}
\end{figure}

\begin{figure}[h]
  \includegraphics[width=\textwidth]{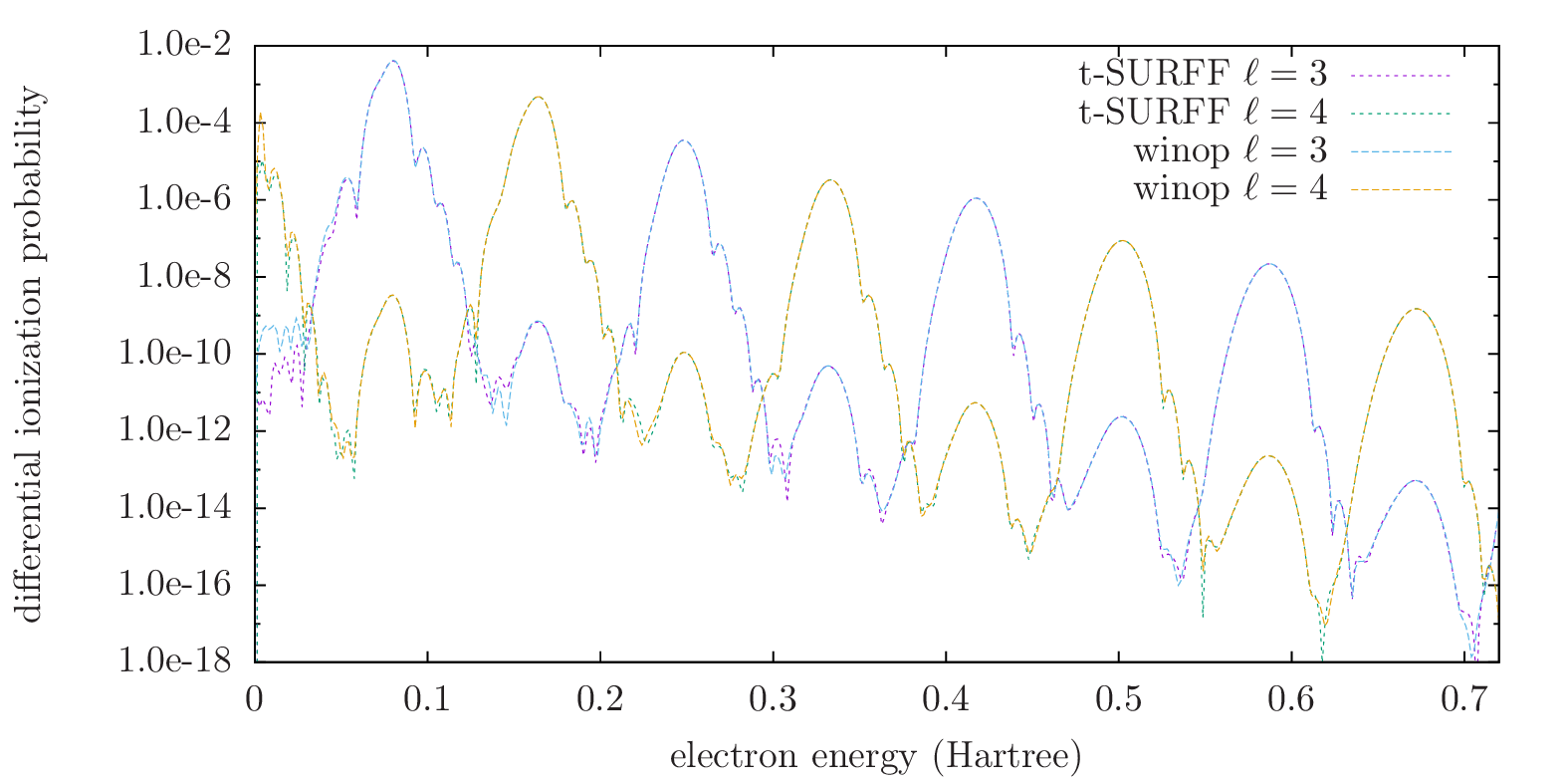}
  \caption{\label{fig:ati-partial} (Color online) Contributions from the $\ell=3$ and $\ell=4$ partial waves to the energy-differential spectrum.}
\end{figure}

In the first example we show that PES calculated by the t-SURFF approximation are in good agreement with spectra calculated using WOM.  
All relevant files are located in the directories \verb!ati-tsurff! and \verb!ati-winop!, respectively. 

In order to ensure that the binding potential vanishes before the t-SURFF boundary a modi\-fied Coulomb potential
\begin{equation}
  \label{eq:binding-potential-co}
  V_{\text{bind}}(r)=
  \begin{cases} \displaystyle
   -\frac{1}{r} & \text{if } r < R_{\text{co}} \\
 \displaystyle  \frac{r- R_{\text{co}}}{R_{\text{co}}^2}-\frac{1}{R_{\text{co}}}   & \text{if } R_{\text{co}} \leq r < 2 R_{\text{co}}\\
0 & r \geq 2 R_{\text{co}}
  \end{cases}  
\end{equation}
with $R_{\text{co}}=25$ (parameter \verb!pot-cutoff! in  \verb!initial.param!) is used. The t-SURFF boundary is at $\RI=100  = 4 R_{\text{co}}$ (parameter \verb!R-tsurff!) to ensure that even highly-excited bound states are negligible for $r>\RI$.

A linearly polarized $n_\mathrm{c}=20$-cycle laser pulse described by the vector potential
\begin{equation}
  \label{eq:vec-potential-34}
  \vecA(t)=\unitvec{z} A_z(t); \quad A_z(t)=\hat{A}\sin^2 \left(\frac{\omega t}{2 n_{\text{c}}}\right) \sin(\omega t+\varphicep)
\end{equation}
with $\omega=0.085$ (wavelength $\lambda=535$\,nm), electric field amplitude $\hat E= \hat A \omega= 0.02387$ (peak intensity $I=2\times 10^{13}\text{ W/cm}^2$), and carrier-envelope-phase $\varphicep=0$ is considered.

For this first example we provide step-by-step directions.
\begin{enumerate}
\item Switch to the directory {\tt qprop-with-tsurff/src/ati-tsurff}. You may give a look to {\tt readme.txt}, {\tt Makefile}, and the \verb!*.param! files.  
\item Type {\tt make} (or first {\tt make clean} and then {\tt make}).
\item Run the imaginary-time propagation by entering {\tt ./hydrogen\_im}. The ground state is quickly reached within the 5000 imaginary-time steps (specified in {\tt hydrogen\_im.cc}).   The executable {\tt hydrogen\_im} generates some output files: the initial wavefunction is stored in {\tt hydrogen\_im-wf\_ini.dat} (real and imaginary part in column 1 and 2, respectively), the final wavefunction in {\tt hydrogen\_im-wf\_fin.dat}, some observables in {\tt hydrogen\_im-observ.dat}, and grid parameters in \verb!hydrogen_im-0.log!. 
\item Run the real-time propagation by entering {\tt ./hydrogen\_re}.  The total number of real time steps 45568 is determined automatically from the sum {\tt long(} $ N_{T_\mathrm{p}}+N_\mathrm{t-SURFF}+1${\tt )} of the pulse duration (in time steps)
\[ N_{T_\mathrm{p}} = \frac{n_\mathrm{c}\, 2\pi/\omega}{\Delta t} = 29567.931\]
and the time
\[ N_\mathrm{t-SURFF} = \frac{\RI/p_{\min}}{\Delta t} = 16000.0 \]
the slowest electron of interest (with momentum $p_{\min}$, assigned to \verb!p-min-tsurff! in \verb!tsurff.param!) takes to arrive at $\RI$ so that it will be captured for the t-SURFF PES. The real-time propagation takes less than 4\,min.\ on our Intel Core i5-3570 desktop computer. The output file {\tt hydrogen\_re-vpot\_z.dat} contains the vector potential $A_z(t)$ (2nd column) vs  time (1st column), the log-file \verb!hydrogen_re.log! grid, time, and laser parameters. The file {\tt hydrogen\_re-obser.dat} contains time, the instantaneous energy expectation value $\bra{\Psi(t)} \hat T + V_\mathrm{bind}(r) \ket{\Psi(t)}$ (where $\hat T$ is the kinetic energy), the projection on the initial state $|\braket{\Psi(0)}{\Psi(t)}|^2$, the total norm on the grid (drops below unity because of the absorbing potential), and the position expectation value $\langle z\rangle=\bra{\Psi(t)} \hat z \ket{\Psi(t)}$.  Initial and final wavefunction are stored in {\tt  hydrogen\_re-wf.dat} as described in the original \qprop{} paper \cite{bauer2006qprop}. The values $1 - $total norm on the grid after the simulation and $1-|\braket{\Psi(0)}{\Psi(t_\mathrm{final})}|^2$ are stored in {\tt hydrogen\_re-yield.dat}. The relevant output files for the subsequent t-SURFF post-processing are \verb!tsurffpsi.raw! and \verb!tsurff-dpsidr.raw!, containing the partial radial wavefunctions and their derivative at $r=\RI$, respectively.
\item  Run the t-SURFF analysis by entering {\tt ./eval-tsurff}. The wall-clock run time should be less than 3 minutes on a state-of-the-art desktop PC. The spectra are stored in {\tt tsurff-partial0.dat} and {\tt tsurff-polar0.dat}. In this example, we focus on the total energy spectrum  \eqref{eq:tsurff-total} and the partial contributions $\sqrt{2\epsilon}\big\vert \abar_{\text{I, t-SURFF},\ell m}(\sqrt{2\epsilon}) \big\vert^2 $ to it so that  only  {\tt tsurff-partial0.dat} is needed. The columns in this file contain (in the case of linear polarization)
energy $\epsilon$, momentum $k=\sqrt{2\epsilon}$,   partial contributions $k\big\vert \abar_{\text{I, t-SURFF},\ell=0, m_0}(k) \big\vert^2, \ldots , k\big\vert \abar_{\text{I, t-SURFF},\ell=L_{\max}-1, m_0}(k) \big\vert^2$, total spectrum ${\diff P_{\text{I, t-SURFF}}(\epsilon)}/{\diff\epsilon}$. Hence, energy is in column 1 and the total spectrum in column $L_{\max}+3$. 

The t-SURFF analysis may be executed in parallel, as explained in {\tt readme.txt}. 
\item The {\tt gnuplot} script  {\tt plot-total-spectrum.gp} generates the graphics file {\tt total-spectrum.png} containing the total t-SURFF PES shown in Fig.~\ref{fig:winop-vs-tsurff}. For comparison, the script also includes the WOM result (to be calculated next), if present.
\item The {\tt gnuplot} script  {\tt plot-partial-spectra.gp} generates {\tt partial-spectrum.png} with the partial t-SURFF PES for $\ell=3$ and $4$, shown in Fig.~\ref{fig:ati-partial}. Again, the script includes the analogous WOM results, if present.
\end{enumerate}

Now we generate the corresponding results using WOM.
\begin{enumerate}
\item Switch to the directory {\tt qprop-with-tsurff/src/ati-winop}. The parameter file {\tt initial.param} is identical to the one for t-SURFF. However, in {\tt propagate.param} the radial grid size for real-time propagation is now explicitly specified ({\tt R-max double 4000.0}, total radial grid size {\tt R-max}$+${\tt imag-width}) whereas in t-SURFF it is calculated automatically as {\tt imag-width}$+\RI+\hat E/\omega^2$ (which is only 253 for this example). There is another parameter file, {\tt winop.param}, discussed below. 
\item Type {\tt make} (or first {\tt make clean} and then {\tt make}).
\item Run the imaginary-time propagation by entering {\tt ./hydrogen\_im}. 
\item Run the real-time propagation by entering {\tt ./hydrogen\_re}. The run time is much longer now ($\simeq$ 38 min.\ on our desktop computers) because of the by a factor 16 bigger grid, which more than obliterates the advantage due to the smaller number of real time steps {\tt long(} $ N_{T_\mathrm{p}}+1${\tt )}$=29568$. The \verb!hydrogen_re*.dat! output files are structured as in {\tt ati-tsurff}. For instance, in \verb!hydrogen_re-obser.dat!, column 4, it is seen that the norm on the larger grid stays unity now whereas in {\tt ati-tsurff} it drops down because the part of the wavefunction representing ionization is absorbed soon after it passed the t-SURFF boundary. 
\item  Run the WOM analysis by entering {\tt ./winop} (takes less than 3 minutes). The parameters in {\tt winop.param} determine that the PES are calculated for {\tt num-energy} values between {\tt energy-min} and {\tt energy-max}. Moreover, more radial grid points may be used for the WOM analysis in order to have a better representation of the continuum (parameter {\tt winop-radial-grid-size} is set to {\tt 50000} in the example). If the radial grid for WOM is too small discrete, ``spherical box'' states are visible. The result in {\tt spectrum\_0.dat} has the same structure as {\tt tsurff-partial0.dat} above. 
\item The WOM PES are included in the output generated by the  {\tt gnuplot} scripts  {\tt plot-total-spectrum.gp} and {\tt plot-partial-spectra.gp} in directory {\tt ati-tsurff}, i.e., Figs.~\ref{fig:winop-vs-tsurff} and \ref{fig:ati-partial}.
\end{enumerate}

Figure~\ref{fig:winop-vs-tsurff} shows the energy-resolved spectra for electron emission  calculated by the t-SURFF and window operator method respectively.
The normalized WOM result  and the corresponding $|\abar_{\text{I, t-SURFF}}(k)|^2 k$ with $k=\sqrt{2 E}$ from t-SURFF are plotted.
The agreement is very good; the results only differ for low energies, as expected.

Both with WOM and t-SURFF the contributions of partial waves of angular momentum index $\ell$ to the energy-differential ionization probability can be computed.
Figure~\ref{fig:ati-partial} shows a comparison of $\big\vert \abar_{\text{I, t-SURFF},\ell0}(k) \big\vert^2 k \vert_{k=\sqrt{2\epsilon}}$ calculated by t-SURFF 
and $\big\vert a_{\text{winop},\ell0}(\epsilon) \big\vert^2$ from WOM (see~\eqref{eq:tsurff-total} and~\eqref{eq:window-prob}, respectively) for the partial contributions $\ell=3$ and $\ell=4$.

\subsection{Ionization of hydrogen in a strong linearly polarized laser field}\label{sec:second-example}

\begin{figure}[h]
 \includegraphics[width=1.0\textwidth]{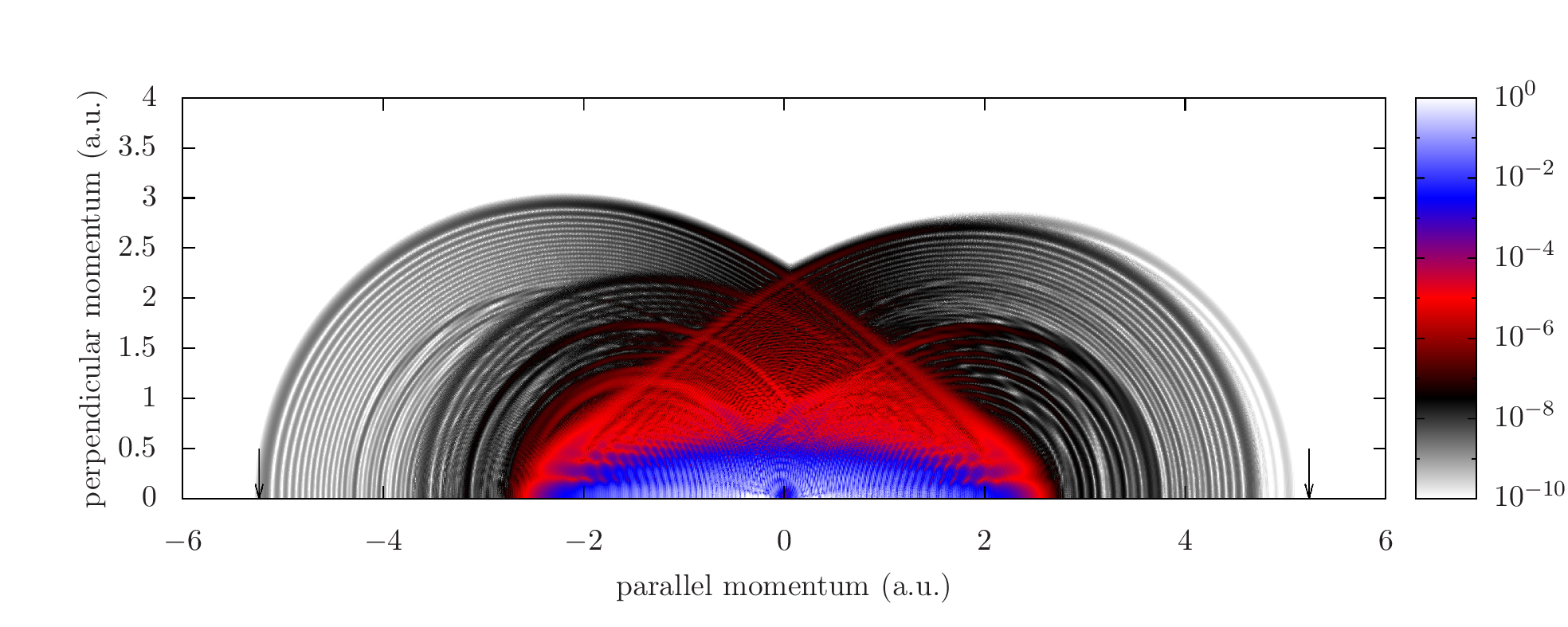}
 \caption{\label{fig:mom-resolved-spectrum-2000-34} (Color online) Momentum-resolved electron spectrum. 
Laser parameters: $\lambda=2000$ nm, $I=10^{14}\text{ W/cm}^2$, $n_{\text{c}}=6$.
A bigger number of angles $N_{\theta_k}=400$ and an extended additional propagation time $T_{\text{t-SURFF}}=2000$ than in the example were used to produce the data for this high-resolution PES.}
\end{figure}

In this example, the momentum-resolved PES shown in Fig.~\ref{fig:mom-resolved-spectrum-2000-34} for a hydrogen atom is calculated for laser parameters
which make the numerical simulations much more demanding than in the previous example. This example is found in the directory \verb!large-clubs!.

The binding potential~\eqref{eq:binding-potential-co} with the cut off radius $R_{\text{co}}=100$ is used. The ground state is obtained after typing {\tt make} and running \verb!./hydrogem_im!, as in the previous example.
Entering \verb!./hydrogen_re! starts the real-time propagation, simulating the interaction with a linearly polarized  $n_{\text{c}}=6$-cycle laser pulse of wavelength $\lambda=2000$~nm, intensity $I=10^{14}\text{ W/cm}^2$, and shape \eqref{eq:vec-potential-34}. It takes about 5 hours on our desktop computer. 
A rough, conservative estimate for the maximal, relevant orbital angular momentum quantum number is $L_{\max} \simeq  (I_{\text{p}}+10 \UP)/\omega\simeq 623$ where $I_{\text{p}}=0.5$ is the ionization potential and $\UP=\hat{A}^2/4\simeq 1.37$ is the ponderomotive potential.
The distance of the t-SURFF boundary $\RI$ should be larger than the classical quiver amplitude $\hat{A}/\omega\simeq 103$ of a free electron in that laser field.
Additionally, wavefunctions of high-lying bound states should be negligible beyond $\RI$, which is ensured if $\RI$ (here 300) is sufficiently larger than  $R_{\text{co}}$ (here 100). The smallest momentum of interest {\tt p-min-tsurff}, determining the post-laser propagation time as explained in the first example, is chosen {\tt 0.5}.
If one is interested in lower-energy regions (to see the low energy structures, for instance \cite{PhysRevX.5.021034}) one should use a smaller value for \verb!p-min-surff!. It may be more efficient to use WOM instead of tuning {\tt p-min-tsurff} down to tiny values.

Next, the momentum-resolved PES calculated by the absolute square of~\eqref{eq:expansion1} is calculated by executing \verb!./eval-tsurff! (or the parallel version, see {\tt readme.txt}). In the interest of a shorter execution time (still 12.4 hours though), a smaller number of angles, a larger \verb!p-min-tsurff! and a larger grid spacing are used  than for the PES shown in Fig.~\ref{fig:mom-resolved-spectrum-2000-34}.

The gnuplot script {\tt plot-polar-spectrum.gp} plots the momentum-resolved PES in the $p_zp_x$-plane from the momentum-angle (i.e., $k$, $\theta_k$) data.     The high-resolution PES in Fig.~\ref{fig:mom-resolved-spectrum-2000-34} beautifully shows many of the textbook features of a strong-field PES in the tunneling regime (the Keldysh parameter is $\gamma=\sqrt{I_{\text{p}}/2\UP}\simeq 0.43<1$): the typical club structure caused by electron rescattering, ``holographic side lobes'' \cite{huismans2011}, and intra-cycle interference \cite{Arbo2010}. Arrows indicate the $p_{\text{max}}=\sqrt{2\times 10 \UP}\simeq 5.2$ cutoff for rescattered electrons along the polarization axis. However, because of the short pulse duration different rescattering clubs belonging to different half laser cycles are visible.

Assume we wanted to obtain the same spectrum with WOM. A conservative estimate for the radial grid size is $R_{\text{winop}}=p_{\text{max}} T_{\text{p}}/2\simeq 4328$. With t-SURFF we have only $R_{\text{t-SURFF}}\simeq 550$.
The advantage of t-SURFF is even more pronounced for simulations of more laser cycles because the computational cost for propagation scales $\sim T_{\text{p}}^2$ for WOM but only $\sim T_{\text{p}}$ for t-SURFF.

\subsection{Hydrogen in a circularly polarized laser field}
We consider ionization by a circularly polarized laser pulse
\begin{equation}
  \label{eq:vec-potential-44}
  \vecA(t)=\unitvec{x} A_x(t)+\unitvec{y} A_y(t), \quad A_x(t)=\hat{A}\sin^2 \left(\frac{\omega t}{2 n_{\text{c}}}\right) \sin\omega t,
  \quad A_y(t)=\hat{A}\sin^2 \left(\frac{\omega t}{2 n_{\text{c}}}\right) \cos \omega t.
\end{equation}
In a circularly polarized few-cycle laser pulse the ionization time is mapped to the electron's angle of escape, constituting a so-called ``attoclock'' \cite{eckle2008attosecond}, which is also the name of the corresponding directory.

We choose $n_{\text{c}}=2$, $\omega=0.114$ (i.e., $\lambda=400$\,nm), $\hat E = \omega\hat A=0.0533799$ (i.e., $I=10^{14}\text{ W/cm}^2$) in  \verb!propagate.param!. The binding potential~\eqref{eq:binding-potential-co}  with the cutoff  $R_{\text{co}}=25$  is used (see \verb!initial.param!).


\begin{figure}[h]
  \includegraphics[width=1.0\textwidth]{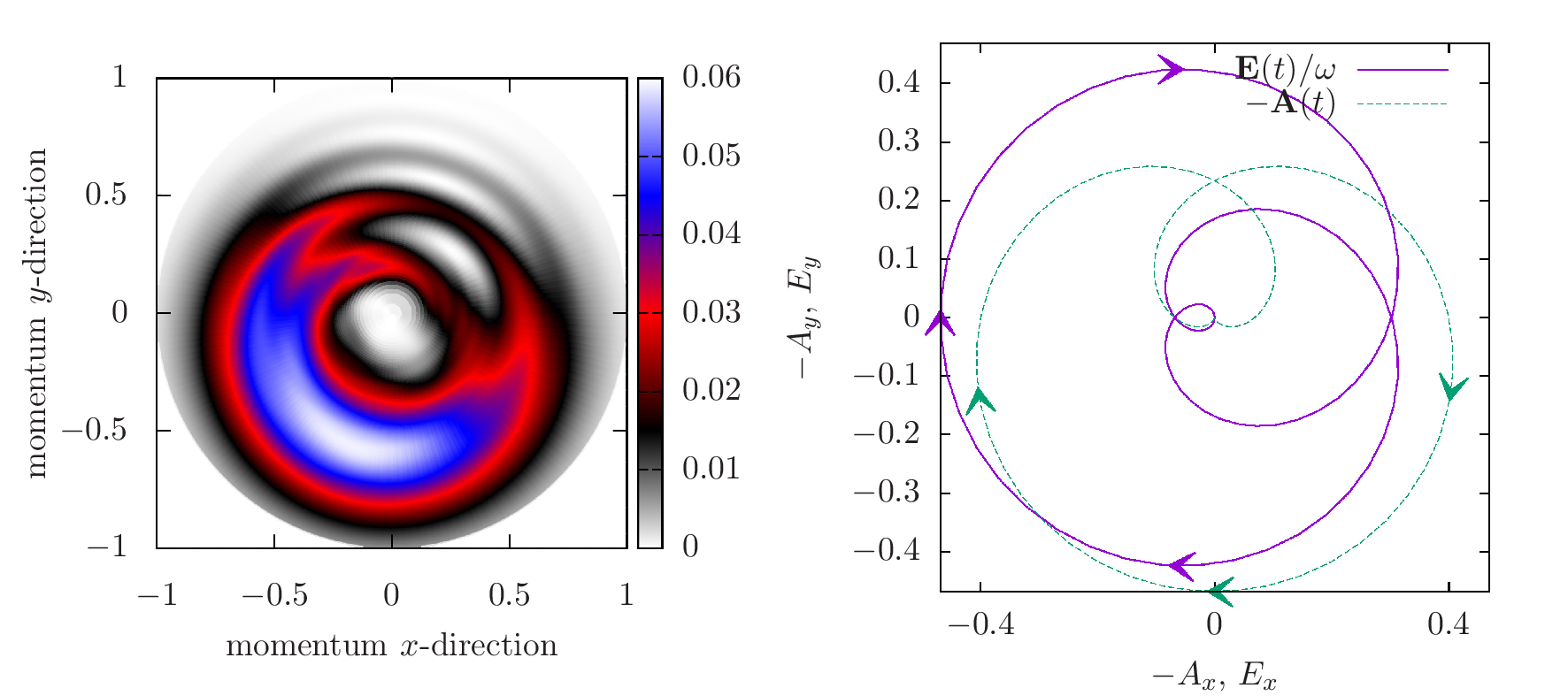}
  \caption{\label{fig:mom-resolved-spectrum-400-44} (Color online) Left: Momentum-resolved ``attoclock'' PES $k\left| a_{\text{I, t-SURFF}}(\veck)\right|^2$ for a two-cycle, circularly polarized laser pulse. Laser parameters: $\lambda=400$ nm, $I = 10^{14}\text{ W/cm}^2$. Right: Electric field and (negative) vector potential.}
\end{figure}

As in the previous examples, after the generation of the ground state via running \verb!./hydrogen_im! the real-time propagation is started by entering \verb!./hydrogen_re!. On our desktop computer this takes 75 minutes.  In the file \verb!hydrogen_re-obser.dat! the columns contain time, field-free energy expectation value $\langle H_0\rangle$, projection on initial state, norm on the grid, and the position expectation values $\langle x\rangle$ and $\langle y\rangle$. 

PES are calculated with \verb!./eval-tsurff! (or \verb!mpirun -np !$n$\verb! eval-tsurff-mpi! for $n$ processes using MPI, see {\tt readme.txt}). The number of $\theta_k$ angles $N_{\theta_k}$   and the number of $\varphi_k$ angles $N_{\varphi_k}$  are defined in {\tt tsurff.param}. Here, in the attoclock example we set $N_{\theta_k}=3$ and $N_{\varphi_k}=50$. The run time for   \verb!./eval-tsurff! is 4.4 hours (and correspondingly faster when processed in parallel). For circular or elliptical polarization in the $xy$-plane the momentum-resolved PES in the $p_xp_y$-plane is most interesting (unless the wavefunction has a nodal plane there). To that end, the bash shell script {\tt select-theta.sh} selects the data for $\theta_k=\pi/2$ from the \verb!tsurff-polar!$i_{\text{p}}$\verb!.dat! file(s) and stores it in \verb!tsurff-polar.dat!. Finally, the gnuplot script \verb!plot-polar-spectrum.gp! plots the PES and generates the graphics file \verb!polar-spectrum.png!.

Figure~\ref{fig:mom-resolved-spectrum-400-44} shows the PES (albeit for higher $N_{\varphi_k}$) whose main features can be explained in simple terms: As the right panel shows, the electric field $\vecE(t)$ peaks in the middle of the pulse, pointing in negative $x$ direction. This implies that the tunneling exit for the electron is at positive $x$ at that most likely emission time. The final drift of the photoelectron according to ``simple man's theory'' (see, e.g., \cite{miloreview,muba2010}) is given by the negative vector potential at the time of emission, pointing in negative $y$ direction. Hence, if the Coulomb interaction between emitted electron and parent ion was negligible, one would expect a maximum probability in the momentum-resolved PES around $p_x=0$ and $p_y=-\sqrt{2\UP}$, where $\UP=\hat A^2/2$ for the vector potential \eqref{eq:vec-potential-44}. However, the Coulomb attraction affects the trajectory of the escaping electron such that it ``swings by'' and accumulates a drift $p_x<0$, explaining why the maximum rotates  clockwise away from this expected position, as seen in the left panel of Fig.~\ref{fig:mom-resolved-spectrum-400-44} \cite{PhysRevA.77.053409}. An additional rotation might be due to a finite tunneling time \cite{ecklescience2008}. An interference pattern is observed in the first quadrant of the momentum plane, which is due to the two trajectories leading to the same final drift momentum, i.e., the crossing of the $-\vecA(t)$ curve in Fig.~\ref{fig:mom-resolved-spectrum-400-44}, right, Coulomb-rotated clockwise away from the  positive $p_y$ axis. TDSE simulations for a similar  setup  where reported in, e.g.,~\cite{martinymadsen2009,Li:15,PhysRevA.93.023425}. Qprop in propagation mode {\tt 44} was used in \cite{PhysRevA.66.053411,PhysRevA.77.053409,Li:15}.

\subsection{Changing the pulse shape}
An example for a pulse shape different from the $\sin^2$-case \eqref{eq:vec-potential-34} is given in the directory {\tt pow-8-sine}. We consider linear polarization, $\vecA(t)=\unitvec{z} A_z(t)$, with
\begin{equation}
  \label{eq:vec-potential-34-sine8}
  A_z(t)=\hat{A}\sin^8 \left(\frac{\omega t}{2 n_{\text{c}}}\right) \sin(\omega t + \varphicep).
\end{equation}
The power-of-eight envelope is sometimes preferable to the sine-square because the spectral decomposition of the laser pulse is closer to realistic, experimental circumstances. The other parameters are kept the same as in the {\tt ati-tsurff} example of Section \ref{sec:first-example}.

The only modifications necessary to implement a ``new'' vector potential are in {\tt class vecpot}, defined  in {\tt potentials.hh} (in the respective example directory). However, apart from the vector potential pulse shape itself, the corresponding time integral, i.e., the excursion $\alpha(t)$, has to be specified there as well. The latter is needed for the t-SURFF post-processing.

After the usual sequence of running {\tt make}, \verb!./hydrogen_im!, \verb!./hydrogen_re!, \verb!./eval-tsurff! (a matter of a few minutes), {\tt gnuplot  plot-total-spectrum.gp} and {\tt gnuplot  plot-partial-spectra.gp} generate the spectra analogous to Fig.~\ref{fig:winop-vs-tsurff} ({\tt total-spectrum.png}) and \ref{fig:ati-partial}  ({\tt partial-spectra.png}). The ATI peaks have less substructure than for the $\sin^2$-envelope. The gnuplot script {\tt plot-polar-spectrum.gp} produces the momentum-resolved PES in {\tt tsurff-mom-res.png}. A typical multiphoton, ATI-like pattern is observed.

\section{Summary}
We incorporated the time-dependent surface flux method (t-SURFF) for the calculation of momentum-resolved photoelectron spectra (PES)  into the \qprop{} package. In that way we facilitate the simulation of momentum-resolved PES up to the fastest relevant electron energies (typically ten times the ponderomotive energy) for laser parameters that were inaccessible with the previous version of \qprop\ based on the window-operator method. In fact, while t-SURFF gets along with  grid sizes of the order of the quiver amplitude, the window operator method requires the full, very delocalized wavefunction at the end of the pulse. Especially for long-wavelengths, high intensities, and many laser cycles t-SURFF is numerically much more efficient than the window operator approach as far as the energetic electrons are concerned. Complementary, the slow electrons (and the bound part of the spectrum) can still be calculated using the window operator since the necessary information is contained in the (non-absorbed) wavefunction on the small grid within the t-SURFF boundary. 

Several examples were provided, whose execution should enable users to adapt \qprop\ to their own problems.

\section*{Acknowledgments}
This work was supported by the SFB 652 of the German Science Foundation (DFG).

\section*{Appendix A}
The TDSE for an electron in a binding potential  $V(\vecr)$ and coupled to an external vector potential in dipole approximation reads
\begin{equation}
  \label{eq:app-tdse}
  \imagi \partial_t \Psi(\vecr, t)
=\left\lbrace \frac{1}{2}\left[\vecp + \vecA(t)\right]^2 
+ V(\vecr) \right\rbrace \Psi(\vecr, t).
\end{equation}
The transformation
\begin{equation}
  \label{eq:app-trafo}
  \Psi(\vecr, t)=\Psi'(\vecr, t)\ \eulere^{-\imagi\int^{t}\diff t' A^2(t')/2}
\end{equation}
yields the TDSE
\begin{equation}
  \label{eq:app-tdse-trafo}
  \imagi \partial_t \Psi'(\vecr, t)
=\left\lbrace \frac{\vecp^2}{2} + \vecA(t)\cdot\vecp 
+ V(\vecr) \right\rbrace \Psi'(\vecr, t)
\end{equation}
without the $A^2(t)$ term. The corresponding Hamiltonian $\hat H = {\vecp^2}/{2} + \vecA(t)\cdot\vecp 
+ V(\vecr)$, with $\vecp=-\imagi \nabla$, is used in (\ref{eq:hamiltonian-3D}).

\section*{References}
\bibliography{qproptsurff}

\end{document}